# Influence des mécanismes dissociés de ludifications sur l'apprentissage en support numérique de la lecture en classe primaire

Raphaël MARCZAK[1,*], Pierre HANNA[2,*], and Christine HANNA[3]

[2]LaBRI, Université de Bordeaux
[1]SCRIME, Université de Bordeaux
[2]Ecole Elémentaire Malartic, Gradignan
[*]prenom.nom@labri.fr

## ABSTRACT

L'introduction des jeux-vidéo sérieux comme soutiens pédagogiques dans le domaine de l'enseignement est un processus de plus en plus démocratisé. En parallèle de cette introduction, la recherche en "game studies" représente une discipline académique très active qui s'intéresse à l'analyse des systèmes vidéoludiques complexes, notamment en ce qui concerne la compréhension de l'expérience des joueurs, et les mécanismes principaux d'immersion. Cet article créé un lien entre l'utilisation en Cours Préparatoire de nouvelles solutions ludo-pédagogiques, et la recherche capitale sur la motivation des joueurs/apprenants, en détaillant une expérience basée sur un jeu développé spécifiquement (nommé QCM), ludifiant par différents mécanismes d'immersion des fichiers d'apprentissage issus de la pédagogie Freinet. L'apport principal pour la recherche de QCM par rapport à des jeux-sérieux plus traditionnels réside dans la dissociation des mécanismes d'immersion afin d'améliorer la compréhension de l'expérience des utilisateurs. Ce jeu contient également un système de métriques d'utilisation, dont l'analyse indique une relative hausse de la motivation des élèves utilisant QCM au lieu des fiches papiers, mais révèle surtout des grandes différences de comportement selon les élèves et les mécanismes de ludification employés.

## 1 Introduction

La place du numérique dans l'éducation fait l'objet de nombreuses études et expériences visant à en mesurer les impacts et les bénéfices éventuels pour l'apprentissage. Parmi les applications rendues possibles, les jeux sérieux suscitent un intérêt grandissant. Les jeux sérieux (serious games) sont des jeux dont l'objectif principal concerne l'acquisition d'une ou plusieurs notions fondamentales[1], rendant ainsi attrayant et motivant un objectif pédagogique précis. Les premiers jeux sérieux datent des années 70[2]. Les domaines d'apprentissage appréhendés sont très nombreux et concernent aussi bien le milieu professionnel que l'enseignement des matières fondamentales pour les enfants.

L'utilisation de jeux sérieux dans l'enseignement est un sujet d'étude important au travers de la mise en place d'expériences menées dans des classes. Ces études montrent un intérêt certain des jeux sérieux pour l'apprentissage, avec cependant des résultats différents selon les matières concernées[3]. Certaines expériences indiquent des retours très positifs[4], comme par exemple l'utilisation raisonnée des tables interactives en classes de Cours Préparatoires (CP)[5].

En lieu et place de "jeux sérieux", le terme de ludification est souvent utilisé. Or, des différences fondamentales existent, et même si la frontière entre jeu sérieux et ludification (ou gamification) est perméable, il souvent difficile de déterminer avec certitude dans quelle catégorie se situe un développement ludo-pédagogique. Andrzej Marczewski a tenté d'expliciter cette différence fondamentale entre ludification et jeu sérieux[6]. Pour lui, un jeu sérieux doit avant tout être une entité basée sur un concept de gameplay: la pédagogie doit être repensée au cœur d'un game-design spécifiquement conçu pour l'acquisition des notions. La ludification, au contraire, prend comme point de départ une pédagogie déjà existante, et l'enveloppe de concepts issus de la ludologie. Nous voyons dans la section 2 de cet article que le projet QCM présenté dans cet article est plus proche de la notion de ludification, que de celle de jeu sérieux.

Les études et expériences existantes contienent de fortes limitations qui proviennent essentiellement du fait qu'une idée de jeu sérieux est testée à chaque fois pour évaluer son game-design global, sans réellement décomposer celui-ci selon ses composantes d'immersions. Pourtant, chaque individu ayant une sensibilité différente aux différents dimensions de gameplay, un seul jeu ne peut pas convenir à tous les enfants d'une même classe. Dans l'étude présentée dans cet article, l'objectif n'est pas de focaliser sur le développement et l'intérêt d'un jeu sérieux particulier, mais d'étudier l'intérêt porté par des élèves face à plusieurs types de mécanismes de ludifications. Il ne s'agit donc pas de proposer de nouveaux jeux sérieux ou de nouveaux mécanismes de ludifications, mais d'étudier leurs impacts éventuels sur la motivation, et également les différences de comportements des individus sur des jeux différents visant le même objectif pédagogique.

Le contexte pédagogique de l'étude décrite dans cet article est la consolidation de l'apprentissage de la lecture en classe de cours préparatoire (CP), avec des enfants âgés de 6 à 7 ans. Des fiches de travail personnel, issues de la pédagogie Freinet, ont été ludifiées selon plusieurs dimensions d'immersion et proposées ainsi aux élèves. Les questions ouvertes par ce travail sont nombreuses : la ludification peut-elle aider à motiver pour le travail en autonomie ? Quels sont les impacts des différents mécanismes de ludification sur les élèves ? Chaque élève a-t-il un comportement propre devant les jeux proposés ? Pour tenter de répondre à ces questions, l'article est découpé en plusieurs sections complémentaires. Dans la section 2 sont présentés le jeu QCM, les mécanismes d'immersions d'intérêt et la ludification des fichiers Freinet de travail personnel pour l'apprentissage de la lecture. Ensuite, dans la section 3, les conditions expérimentales sont détaillées, avant de présenter dans la section 4 les différentes observations recensées. Ces observations permettent d'établir un ensemble de premières conclusions et perspectives détaillées dans la section 5.

## 2 Jeux sérieux pour l'apprentissage de la lecture

L'idée centrale du jeu sérieux QCM (la Quête du Cornichon Masqué) est d'étudier l'impact sur l'efficacité de l'apprentissage des principaux mécanismes de motivations vidéoludiques, tels que théorisés par Gordon Calleja (pour l'immersion,[7]) et Mihaly Csikszentmihalyi (pour le flow,[8],adapté aux jeux-vidéo par Jenova Chen[9]).

Gordon Calleja a notamment défini les grandes dimensions de motivation (conduisant à l'immersion), comme étant :

- *La narration* : le joueur est motivé par l'histoire du jeu. Il veut savoir comment l'histoire évolue, et se termine.

- *Le ludique* : le joueur est motivé par les éléments ludiques du jeu. Il veut essayer d'augmenter son score, de débloquer des succès, ou de compléter des niveaux ou puzzles complexes.

- *L'émotion* : le joueur est immergé dans le jeu grâce à la complexité émotionnelle de celui-ci, à travers des choix artistiques, de l'humour, de la tendresse, ou au contraire des émotions plus sombres, comme la mélancolie ou la peur.

- *Le social* : le joueur est motivé par la possibilité d'adopter des comportements sociaux, que ce soit avec d'autres joueurs, ou d'autres avatars (non joueur).

- *La kinesthésie* : le joueur est immergé dans le jeu à travers la sensation de ne faire qu'un avec son avatar, grâce à des contrôles suffisamment intuitifs pour donner l'impression de maîtriser les déplacements et actions de son avatar avec virtuosité.

- *La stratégie* : le joueur est motivé grâce à sa connaissance approfondi du monde virtuel dans lequel il évolue. Il peut alors y créer des stratégie pour atteindre les objectifs du jeu (connaissance des règles, des lieux, des items, etc).

Gordon Calleja parle également de deux niveaux de motivation, ou d'implication : la micro-implication et la macro-implication. La première contient les raisons qui poussent un joueur à continuer une session de jeu, tandis que la deuxième explicite les causes motivant à jouer à nouveau à un jeu une fois la session terminée.

Une autre recherche essentielle à la bonne compréhension de la motivation dans les jeux-vidéo est celle entourant l'état psychologique du Flow, tel que théorisé par Mihaly Csikszentmihalyi[8]. Le Flow est un état psychologique de bien-être ressenti par les individus lorsqu'ils accomplissent une tâche gratifiante. Initialement observé chez les sportifs ou musiciens arrivant à s'auto-motiver lors du processus (long) d'apprentissage et de perfectionnement, le Flow a également été observé chez les joueurs de jeux-video[9], lorsque le game-design a été pensé pour inclure une progressivité efficace de la difficulté. D'après Csikszentmihalyi, l'état de Flow est atteint lorsqu'une balance optimale est trouvée entre la difficulté d'un objectif et la confiance de l'individu dans ses propres capacités. L'objectif doit être ambitieux, pour motiver l'apprenant, sans être trop complexe, pour ne pas l'angoisser. A ce moment là, lorsque l'objectif est atteint, l'individu gagne en confiance, et peut essayer d'atteindre des objectifs plus complexes, et ainsi de suite. L'une des manifestations principale du Flow est le sentiment de ne pas voir le temps passer. Dans les jeux-vidéo le Flow peut être atteint si la difficulté est progressive : on parle également de courbe de difficulté.

L'objectif principal de QCM est ainsi de pouvoir séparer ces grands axes de motivation afin de pouvoir étudier leurs impacts indépendamment, tout en répondant à la diversité des profils des joueurs. En effet, ce qui va motiver un joueur dans un monde virtuel n'est pas forcement ce qui va en motiver un autre. Par exemple, dans un article sur l'analyse des temps forts musicaux dans le jeu-vidéo Bioshock 2[10], il a été remarqué des divergences de comportement fondamentales, qui se sont explicitées lors d'entretiens avec les joueurs. Sur la figure 1, il a été répertorié le temps mis par neuf joueurs pour attendre des moments clefs de l'histoire du jeu. Le point d'intérêt illustrant les différences de motivation est le temps mis entre le deuxième et le troisième événement. En effet, le deuxième point représente la rencontre avec un boss du jeu (nommé Big Sister) qui s'enfuit à la fin d'un combat. Ce boss est rencontré à nouveau, ce qui est représenté par le troisième point. En majorité, les joueurs mettent entre deux et trois minutes pour rencontrer à nouveau le boss. Mais la figure 1 montre deux joueurs se démarquant par leur comportement : le joueur 1, mettant une minute, et le joueur 3 en mettant dix. Le joueur 1 a expliqué sa motivation "ludique", dans le sens où il désirait finir le jeu le plus vite possible. Une fois le boss enfui, il l'a poursuivi sans relâche pour terminer le combat. Le joueur 3, au contraire, a expliqué que le combat avec le boss ne l'avait pas passionné, et qu'il était soulagé de voir l'ennemi s'enfuir. Il a ainsi pu passer du temps à observer le monde virtuel, et à lire et écouter les différents objets du jeu (livres, bandes sonores) décrivant en détail l'histoire et le passé de la cité dans lequel il évolue. Ce joueur était alors motivé par la dimension "narrative". Les autres joueurs avaient quant à eux une approche hybride du jeu.

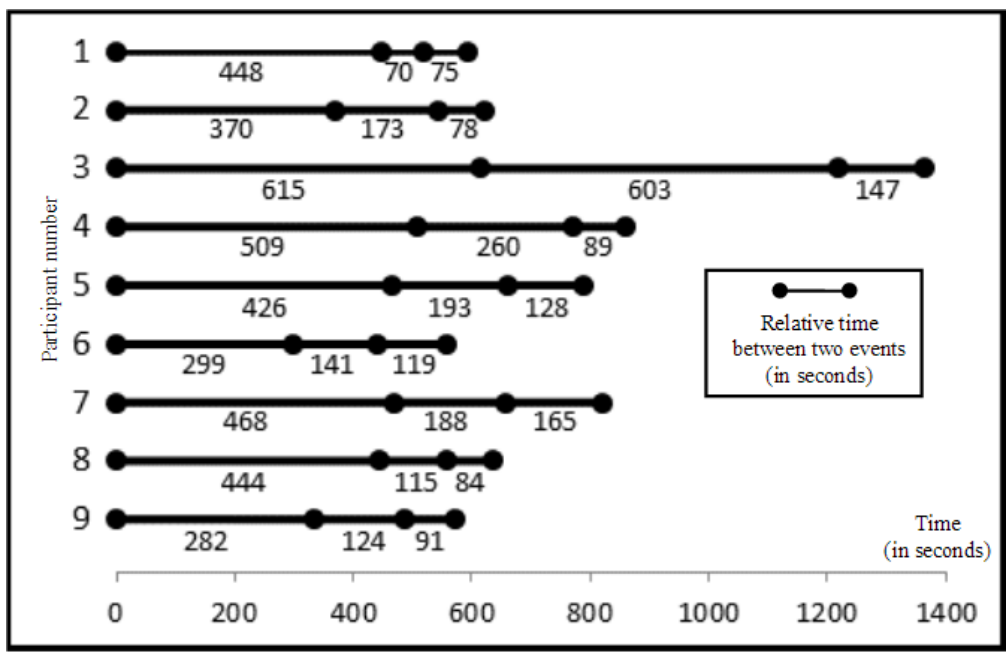

**Figure 1.** Temps mis par des joueurs de Bioshock2 entre quatre temps forts du jeu : premier pouvoir acquis, première rencontre avec le "boss", deuxième rencontre avec le "boss", vision extérieure de la ville.

Au sein des jeux-vidéo (sérieux ou traditionnels), ces dimensions (ludiques, narratives, sociales, émotionnelles, stratégiques et kinesthésiques) co-existent, à des degrés différents, et permettent de créer un monde virtuel complexe dans lequel un joueur peut avoir envie de s'immerger. Mais il est alors difficile de réellement déterminer ce qui motive un élève interagissant avec un jeu pédagogique, et de proposer des expériences personnalisées à même de répondre à la diversité des profils de joueur. Avec la Quête du Cornichon Masqué (QCM), nous avons essayé de proposer un jeu de lecture basé sur un même objectif pédagogique, mais avec des expériences différentes centrées sur des mécanismes de motivation bien séparés.

La figure 2 illustre l'architecture choisie pour le développement de QCM, avec la séparation souhaitée des dimensions pour la réalisation d'expérimentations distinctes. Pour ne pas influencer de manière inappropriée la pédagogie mise en œuvre par Freinet, les fiches d'apprentissages de la lecture restent inaltérée dans QCM : ce sont des versions numérisées de l'existant sur papier. Tout au plus la différence se fait par le biais d'une interaction "clic/touché" au lieu d'une action manuscrite. Ainsi, l'objectif pédagogique reste préservé. En revanche, différentes surcouches vidéoludiques ont été implémentées, et permettent d'adjoindre des dimensions de motivation sans contrefaire la pédagogie initiale. Parmi les dimensions fondamentales présentées précédemment, nous avons choisi d'en extraire trois, la narration, le ludique et l'émotion, en intégrant également la motivation issue du flow (les autres dimensions sont en cours d'implémentation). La narration et le ludique représentent ainsi deux expériences ludo-pédagogiques autonomes, qui font appel, au sein de leurs gameplay, aux fiches de lecture. Par exemple, il faut répondre à une fiche pour faire avancer l'histoire - narration, ou il faut répondre juste pour récupérer une clef - ludique. L'émotion, elle, se situe au sein du menu de choix du jeu, et au sein de chaque jeu. En effet, les jeux et le menu se situent tous dans une ambiance *médiéval/fantastique*, jouant un rôle de *conteur* pour les élèves. Enfin, le flow se situe au cœur de la difficulté progressive donnée aux fiches d'apprentissage de la lecture, et dans une moindre mesure à l'évolution du joueur au sein des jeux narratifs (histoire qui se complexifie) et ludiques (actions à réaliser pour passer aux niveaux supérieurs plus complexes). Ces choix et développements vont être détaillés dans les paragraphes suivants.

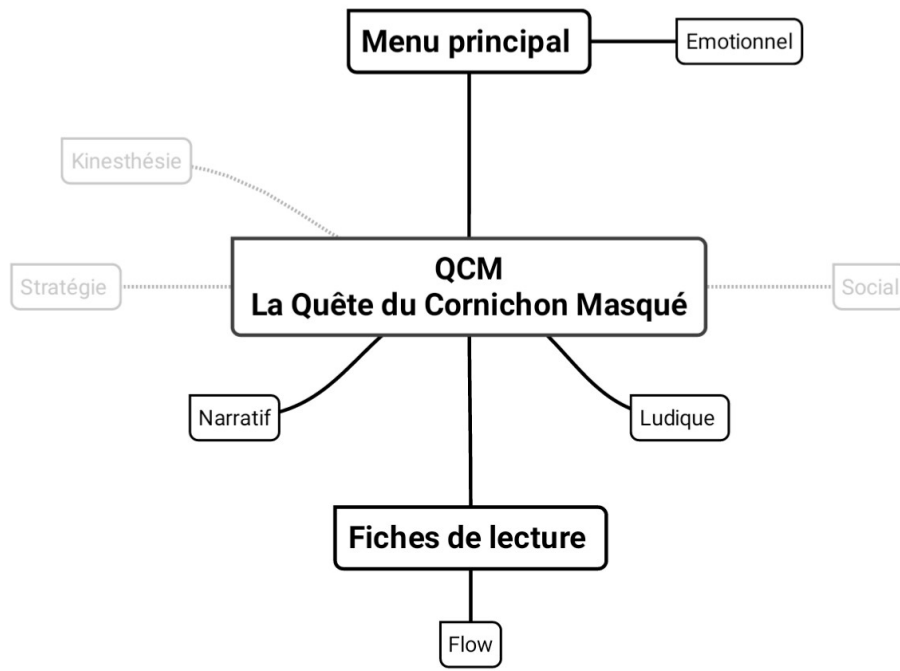

**Figure 2.** Schéma présentant l'architecture générale de QCM, avec les dimensions de motivation associées.

Ainsi, cette section suivante est centrée sur la présentation des fiches pédagogiques liées à l'apprentissage en autonomie de la lecture avec leurs objectifs de travail, puis l'application QCM et ses deux jeux principaux reposant sur ces fiches, en insistant sur les mécanismes de ludification employés.

### 2.1 Pédagogie Freinet et fiches de travail personnel

La pédagogie Freinet[11] est une pédagogie reposant sur l'expression libre et proposant à l'enfant (apprenant) des sessions de courtes durées (20 à 25 minutes par jour pour les CP, jusqu'à 45 minutes dans les classes de CM1 et CM2) de travail personnalisé en autonomie. Des supports spécifiques à ce travail en autonomie ont ainsi été créés autour des différentes matières enseignées en école primaire : Mathématiques, Orthographe, Français, Sciences, etc. Les fichiers auto-correctifs proposés par l'éditeur PEMF & Cie [1] font partie des outils fréquemment utilisés en pédagogie Freinet. Pour constituer la base pédagogique de QCM, nous avons choisi les fichiers de lecture adaptés aux classes de Cours Préparatoire, en cycle II. Il existe 4 niveaux de fichiers, contenant 48 fiches chacun, et ordonnées par difficulté croissante (le 4ème fichier étant proposé pour une liaison CP/CE1). L'objectif pédagogique, tel que spécifié par l'éditeur, est principalement de donner un sens aux éléments écrits et de développer des stratégies de lecture de plus en plus expertes.

Puisque les expérimentations concernent une classe de Cours Préparatoire de fin d'année, nous avons choisi, sur les conseils de l'enseignante de la classe, de considérer les fichiers de niveaux 2 et 3. La figure 3 donne un exemple des fiches de lecture utilisées. Une fiche est constituée d'une première partie, au recto, comportant une illustration et une phrase (ou des mots) s'y rapportant, avec éventuellement des mots partiellement masqués ou manquants. La deuxième partie, au verso, propose une autre situation illustrée en lien avec celle du recto, avec plusieurs affirmations sous forme de texte. L'enfant doit choisir parmi ces propositions la phrase ou le mot correspondant à la situation. Le choix s'effectue en sélectionnant le symbole (carré bleu, triangle rouge, etc) présent devant chaque proposition. Le nombre de propositions peut varier, entre 3 et 6 dans les fichiers sélectionnés.

L'organisation habituelle d'une séance de travail en autonomie personnalisée permet à l'enfant de prendre une fiche du fichier sur lequel il travaille, d'y réfléchir, d'entourer la réponse correspondant sur sa feuille-bilan reprenant l'ensemble des 48 fiches du fichier, puis de faire corriger la feuille-bilan à l'enseignant, une fois celle-ci entièrement remplie. L'enseignant indique alors simplement les erreurs au moyen d'un code couleur.S'il a commis une erreur, l'enfant est invité à reprendre la ou les fiche(s) concernée(s), et à y réfléchir de nouveau pour se corriger. Les jeux étudiés dans cet article ont été développés avec le soucis de respecter cette pratique : après chaque réponse donné par l'enfant, une correction simple du type "juste/faux" est présentée sans plus de détails. En cas de réponse incorrecte, la fiche sera alors présentée de nouveau à l'enfant, dans une session future.

---

[1] http://www.pemf.fr/pemfetcie/

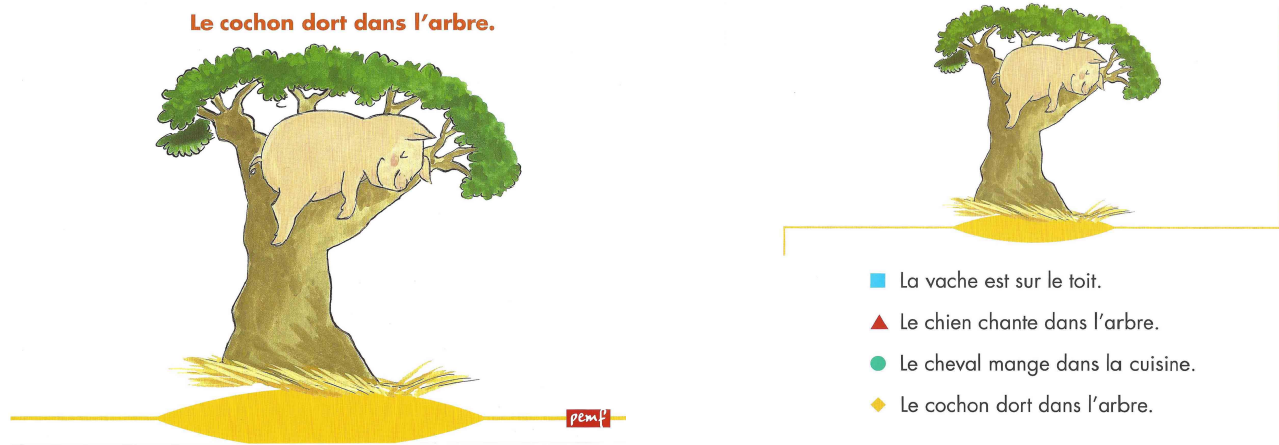

**Figure 3.** Exemple d'une fiche de lecture destinée au travail personnel en autonomie en niveau de Classe Préparatoire (édition PEMF&Cie).

Ces fiches d'apprentissage de la lecture ont été numérisées sous forme d'images, pour permettre de servir de base aux jeux sérieux proposés lors des expérimentations en classe. La pertinence pédagogique de ces fiches n'est plus à démontrer. Leur simplicité et leur objectif de travail en autonomie appuient également ce choix. Ces fiches numérisées sont volontairement présentées de la manière la plus neutre possible, de façon à être le plus proche possible de la présentation des fiches au format papier. Ainsi l'influence des différentes dimensions de ludifications ne perturbe pas directement la pédagogie Freinet, et la comparaison entre le travail sur les tablettes numériques et sur les fiches au format papier est rendue possible.

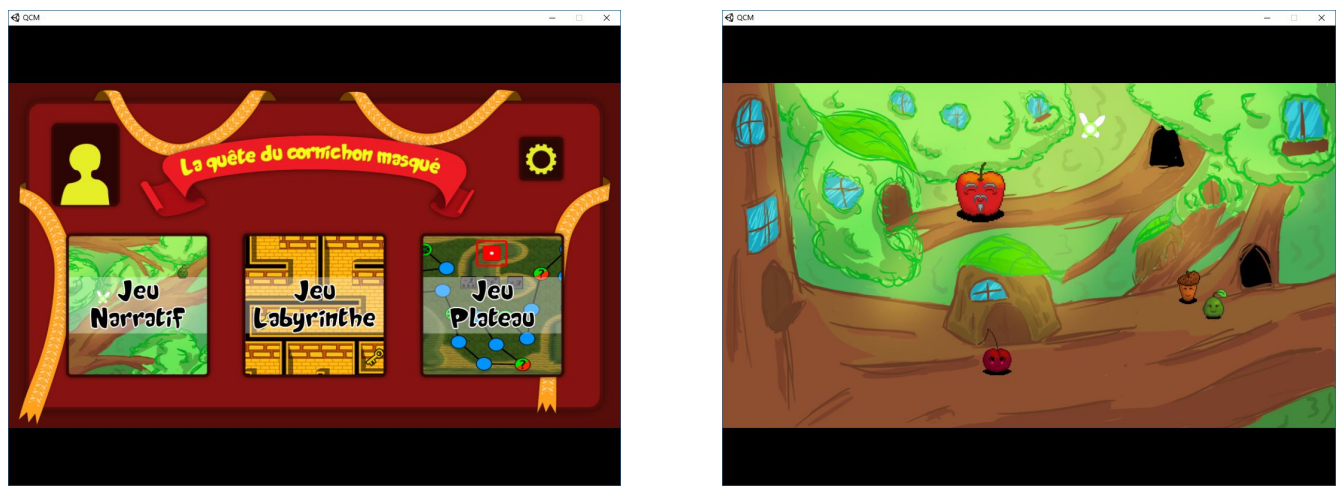

**Figure 4.** Jeux expérimentés en classe: à gauche, le menu principal, à droite, un écran du jeu narratif.

### 2.2 Dimension narrative

Le premier jeu développé au sein de QCM repose essentiellement sur la dimension narrative. Un écran du jeu narratif est présenté sur la figure 4. Cette motivation narrative, telle que formalisée par Gordon Calleja[7], provient de l'intérêt d'un joueur pour la structure narrative d'un contenu vidéoludique. Il s'agit, du point de vue de la micro-implication, de la motivation issue de la curiosité d'un joueur envers le déroulement immédiat de l'histoire du jeu. Si l'on prend l'exemple de la lecture d'un roman, il s'agirait de la motivation permettant à un lecteur d'enchaîner les paragraphes jusqu'à la fin d'un chapitre.

Du point de vue de la macro-implication, la motivation narrative est celle qui va pousser un joueur à revenir vers le jeu pour en atteindre le dénouement final. Si l'on prend à nouveau l'exemple de la lecture, la macro-implication narrative concerne le

**5/18**

moment où le lecteur arrête sa lecture à la fin d'un chapitre, marque la page avec un marque-page, et décide de fermer le livre pour le reprendre ultérieurement.

Gordon Calleja[7] approfondit son analyse de l'immersion narrative en incluant également le joueur au sein de celle-ci. Puisque la nature intrinsèque du média vidéo-ludique est d'être activée par un joueur[12], il est possible pour ce même joueur de "se raconter" en train d'interagir avec un système de jeu. Il peut narrer à son entourage les actions qu'il a choisi de réaliser, et les raisons les justifiant. Ainsi, il est important de penser la narration d'un jeu comme étant à branchements multiples, afin de permettre à un joueur d'avoir le sentiment de contrôler le déroulement narratif du jeu, et ainsi se créer la narration propre de son expérience.

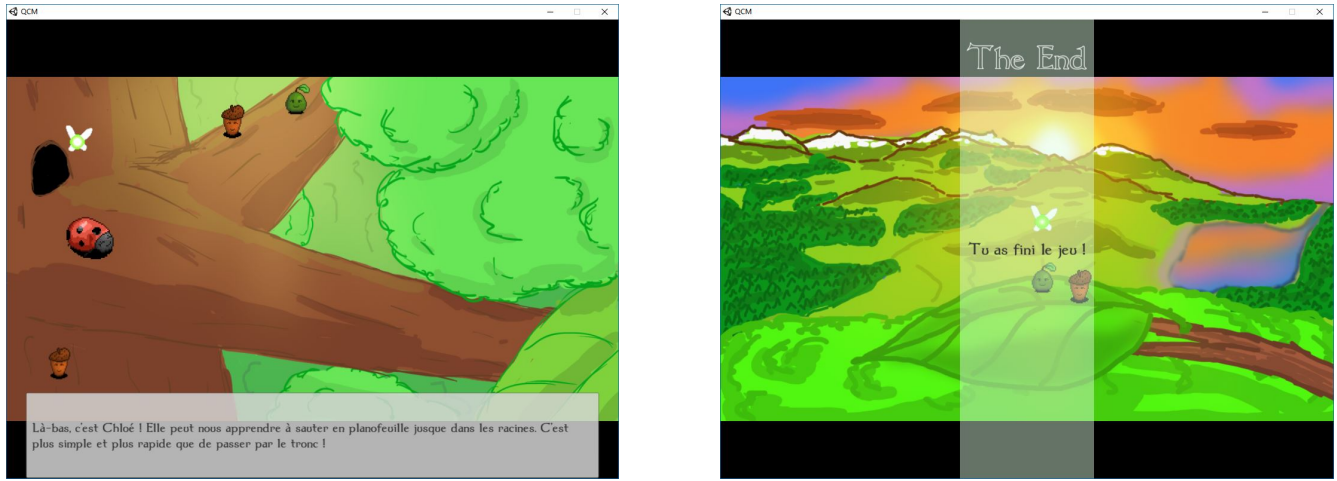

**Figure 5.** Jeu narratif : à gauche, capture d'écran d'une séance de jeu avec du texte, à droite, scène finale.

L'expérience narrative proposée au sein de QCM répond aux problématiques soulevées dans les paragraphes précédents, afin d'optimiser l'engagement du joueur envers le système de jeu. L'histoire est celle d'une petite graine, perdue sur un arbre, et qui ne sait pas encore parler. Elle rencontre différents personnages amicaux, prêts à l'aider dans sa quête sur ses origines, le long de sept scènes dessinées à la main[2]. Le dessin "à la main" permet, du point de vue engagement émotionnel, de se rapprocher du style artistique des enfants de cet âge.

Lorsque l'enfant interagit avec un personnage, celui-ci lui parle. L'enfant entend le dialogue (voix enregistrée), synchronisé avec un texte défilant. Il est ainsi possible pour l'enfant de faire le lien entre ce qu'il entend et la transcription attenante, soit de manière autonome pour les lecteurs "experts", soit avec l'aide de l'adulte présent pour les autres. Le personnage lui propose une quête, et une fiche de lecture apparaît alors. Si l'élève répond correctement, la quête est réussie, et l'histoire suit un embranchement positif. Si l'élève se trompe, la quête n'est pas accomplie, et l'histoire aura un dénouement plus modeste. Dans tous les cas, l'histoire se termine sur une note positive, le but étant de ne pas frustrer inutilement un joueur ayant encore à ce stade des difficultés à lire, par rapport à d'autres camarades. A la fin du jeu, l'avatar du joueur, et les personnages rencontrés, se retrouvent sur la cime de l'arbre (voir la figure 5), et racontent à la graine son futur, qui sera de continuer à peupler la forêt.

Ce jeu engage donc l'élève par la narration, grâce aux mécanismes suivants :

- Une histoire simple, mais immersive, permettant à l'élève d'être motivé pour atteindre la scène suivante (micro-motivation);

- Une histoire à branchements permettant à l'élève de se créer sa propre narration, et d'en discuter avec ses amis;

- Un visuel enfantin pour connecter avec le joueur sur le plan de l'engagement émotionnel;

- Un univers sonore simple et reposant, pour que le joueur se sente à l'aise, et dans de bonnes conditions pour lire et répondre aux questions;

---

[2]graphismes : Henri Toussaint

- Lorsque l'élève se trompe, le dialogue indique clairement que la quête reste inachevée, mais l'histoire continue sur une note positive, afin de ne pas frustrer les élèves ayant encore des difficultés de lecture;
- Chaque dialogue est sous-titré, pour permettre à l'enfant de faire le lien entre les écrits et leurs correspondances parlées, pour certains de manière autonome et pour d'autres avec l'aide de l'adulte.

### 2.3 Dimension ludique

Le deuxième jeu implémenté repose sur l'engagement ludique[7]. Il représente le besoin, pour un joueur, de dénouer un puzzle, de terminer un ou plusieurs objectifs, et/ou d'atteindre un score élevé. Cette dimension est essentielle, et est partie intégrante de tout jeu (vidéo ou non). Même pour le jeu basé sur la dimension narrative, l'envie de trouver la "bonne" fin peut être perçue comme une dimension ludique. Pour limiter l'influence du ludique dans le jeu précédent, nous avons ceci dit évité toute mention d'un score, d'un temps, ou d'objectifs à débloquer.

Tout comme la dimension narrative, l'engagement ludique peut être *micro* ou *macro*. Dans la phase dite *micro*, le joueur essaie de terminer le puzzle ou l'objectif actuel. L'idée est de ne pas quitter le jeu *en cours de route*. Il faut donc des niveaux, des objectifs, ou des phases de jeu bien distinctes, pour que le joueur puisse identifier les instants où un défi est réussi. Au niveau *macro*, le joueur souhaite améliorer son score ou débloquer des succès, ou bien souhaite rejouer parce que les niveaux comportent une propriété de rejouabilité (c'est à dire que le joueur peut les retenter sans s'ennuyer). Gordon Calleja explique également qu'une des forces de l'engagement ludique réside dans la capacité d'offrir des mécanismes d'auto-défi pour les joueurs. Cela signifie que les joueurs peuvent créer leurs propres défis, voir même les proposer aux autres. Par exemple, les jeux Super Mario Bros contiennent des chronomètres. Lorsque celui-ci atteint zéro, le joueur perd. Mais la présence de ce chronomètre permet également aux joueurs de se lancer des défis, du type "je veux finir le niveau en moins de $x$ secondes" ou de dire à ses amis "j'ai fini le niveau en $y$ secondes, peux-tu faire mieux ?".

Le jeu ludique, proposé par QCM, s'inscrit dans les propriétés énumérées précédemment. Il s'agit tout d'abord d'un jeu de labyrinthe, clairement identifiable par des enfants comme un puzzle dont ils maîtrisent les règles : trouver la sortie sans traverser les murs. Pour sortir du labyrinthe, il faut récupérer des clefs (une au niveau un, deux au niveau deux, etc), qui y sont disséminées. Une fois la clef trouvée, une question du fichier de lecture est présentée à l'élève. S'il répond correctement, il récupère la clef. S'il se trompe, la clef est téléportée ailleurs dans le labyrinthe.

Plusieurs choix de game design soutiennent la dimension ludique en plus du l'aspect puzzle/labyrinthe, ludique par nature:

- tout d'abord, un chronomètre est présent sur l'interface du labyrinthe. Celui-ci est ascendant, afin de ne pas stresser le joueur qui pourrait sinon répondre "trop vite et faux" au lieu de "tranquillement et pertinemment". En revanche, la présence de ce chronomètre offre une dimension ludique, permettant aux joueurs (aidés de l'enseignant) de se comparer aux autres élèves. A la fin des cinq niveaux du jeu, un récapitulatif (score) est affiché, niveau par niveau.
- en plus des clefs, des petits personnages peuvent apparaître (comme par exemple un cornichon malicieux). Le joueur peut les rencontrer. Il doit alors répondre à une question, pour "débloquer" ce personnage. Ceci fait appel à la motivation de collection (réussir à avoir tous les personnages), liée également à la notion de succès.
- les niveaux sont également générés procéduralement. Cela signifie qu'il est impossible de prévoir la forme du labyrinthe en amont. Cela signifie surtout que le jeu a une haute valeur de rejouabilité, puisque chaque niveau, aléatoire, sera différent du précédent. Ainsi un joueur ayant fini les cinq niveaux peut y rejouer, et devoir gérer un défi différent.
- enfin, le style graphique de chaque labyrinthe évolue selon les niveaux (murs, bibliothèques, etc), ainsi que l'ambiance musicale. La musique est au début uniquement composée de la basse, puis la batterie se rajoute, et enfin les instruments de premiers plans. Ainsi, plus le joueur avance, plus il est "récompensé" par une musique fournie. C'est une motivation ludique plus inconsciente, mais avant tout un signal fort indiquant bien au joueur qu'il progresse.

Il est important de préciser qu'aucune narration n'est proposée au joueur. Les niveaux s'enchaînent sans transition, autre qu'un changement graphique, sonore et de difficulté. Ainsi, la dimension narrative a peu d'influence sur ce jeu, et la motivation peut être testée indépendamment.

### 2.4 Menu de choix

L'immersion au sein d'un jeu-vidéo commence dès le menu principal. Parfois considéré comme peu important, le menu principal représente pourtant le point d'entrée du joueur dans l'univers virtuel. Il permet, en quelques secondes, de faire entrer le joueur dans le cercle magique[13], c'est à dire d'intégrer le fait que ses actions auront un impact au sein d'un monde aux règles clairement définies, avec des conséquences prévisibles par ces mêmes règles.

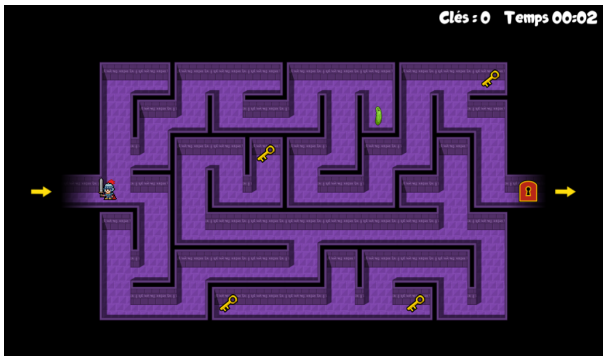 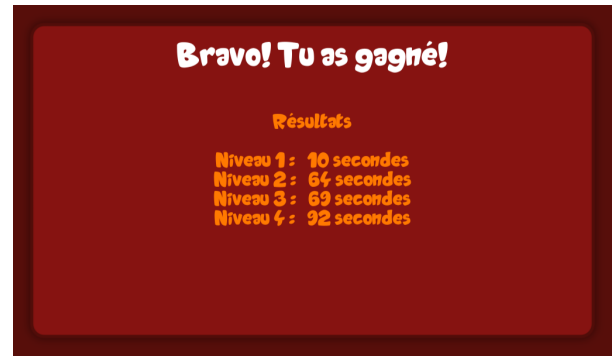

**Figure 6.** Jeu ludique: à gauche, capture d'écran d'une séance de jeu, à droite, fin du jeu avec affichage des résultats.

Les deux jeux présentés précédemment ont une cohérence spatiale et temporelle. Le jeu ludique se passe dans un monde médiéval ; et le jeu narratif est basé sur un système de quête, dans un univers dépourvu de technologie. Ces deux jeux peuvent donc être classifiés comme faisant partie d'univers fantastiques. Afin de respecter cette cohérence, le menu de choix (voir la figure 4) est également d'inspiration fantastique/médiévale. Ainsi, il n'y a pas d'interruption de l'immersion pour le joueur, entre le menu et le jeu qu'il a choisi. Ce choix permet donc de ne pas influencer l'engagement émotionnel et spatial du joueur envers le système vidéoludique.

## 3 Protocole expérimental

Les jeux au sein de QCM, basés sur les fichiers d'apprentissage en autonomie de la lecture, ont été expérimentés en classe. La classe concernée est la classe de Cours Préparatoire de l'école élémentaire Malartic à Gradignan (Gironde, France). La classe était composée de 23 élèves, entre 6 et 7 ans, dont 8 filles et 15 garçons. Deux groupes de 11 et 12 enfants ont été constitués. La répartition des élèves s'est faite en accord avec l'enseignante pour obtenir deux groupes comparables du point de vue des niveaux de lecture de chacun. L'expérimentation a duré 2 semaines en Juin 2016. Les jeux ont été installés sur des tablettes numériques, format 10 pouces, équipées de casques. Chaque enfant avait sa tablette dédiée, labellisée à son nom, permettant alors de récolter les informations relatives (et personnelles) à l'utilisation. Il est à noter que les élèves de cette classe n'avaient jamais utilisé de tablettes numériques en classe, mais avaient parfois accès à des jeux sérieux sur deux ordinateurs, en utilisation libre lors de séances de travail en autonomie. Les photographies de la figure 7 illustrent l'environnement des expérimentations. Le groupe d'élèves travaillant sur les fiches au format papier se trouvait dans la classe, alors que le groupe travaillant sur les tablettes était dans la salle-atelier attenante à la classe.

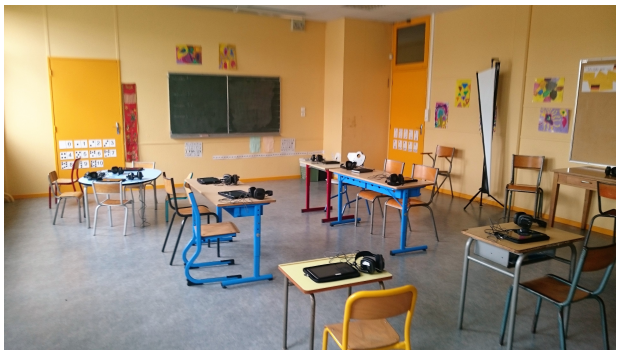 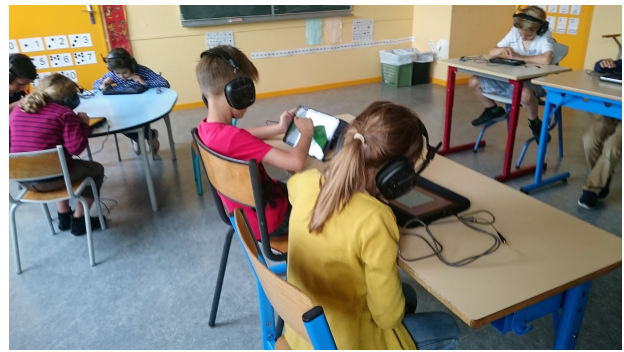

**Figure 7.** Expériences en classe : à gauche, la salle atelier, attenante à la classe, dédiée à l'utilisation de tablettes; à droite, un groupe d'enfants manipulant les jeux sérieux de QCM.

Les deux semaines d'expérimentations ont été scindées en deux, chaque semaine étant dédiée à un des deux groupes d'enfants. Nous notons ces deux groupes *A* et *B* pour la suite. Préalablement, l'enseignante a présenté collectivement aux enfants, par vidéoprojection, les différents jeux proposés, ainsi que le protocole expérimental et la composition des groupes. Chaque semaine débute par une séance dédiée le lundi après-midi : pendant qu'un groupe utilise les jeux sérieux, l'autre groupe travaille sur les fiches de lecture au format papier. Les deux groupes travaillent en autonomie, l'enseignante corrigeant les fiches au format papier, un des co-auteur assistant les enfants sur tablettes en cas de problème technique. Une durée de séance de 20 à 30 minutes a été envisagée en concertation avec l'enseignante. A la suite de ces deux séances (une en début de chaque semaine), un échange a eu lieu avec les enfants pour recueillir leurs avis.

Les choix des fiches de travail personnel consacrées à la lecture, choisies pour la base des jeux sérieux, sont adaptés : la première semaine, tous les enfants ont accès aux fiches (format papier ou numérique) de niveau 2. Lors de la deuxième semaine, les enfants qui ont terminé le ficher de niveau 2 se voient proposer les fiches de niveau 3, alors que ceux qui n'ont pas terminé se voient d'abord proposer les fiches de niveau 2 qu'ils n'ont pas traitées ou réussies.

Au cours de la première séance, les deux jeux ludique (jeu labyrinthe) et narratif (jeu narratif) étaient à disposition, à la volonté des enfants. Le jeu de plateau présenté dans le menu de la figure 4 n'a été rendu accessible que pour les autres séances de travail en autonomie, car ce jeu implique plusieurs joueurs. En raison de difficultés sur les relevés d'utilisation pour ce jeu (utilisation collective), l'étude sur l'utilisation de ce jeu n'est pas détaillée ici, mais le sera dans le cadre d'un projet de recherche futur.

A la suite de cette première séance, les enfants du groupe qui avait utilisé les tablettes numériques pouvaient reprendre les tablettes sur des périodes d'autonomie au cours du reste de la semaine. Ces périodes d'autonomie dépendent du temps restant après des travaux collectifs, et peuvent donc varier selon les enfants. De plus, les enfants avaient le choix entre plusieurs activités en plus de l'utilisation des tablettes : jouer à des jeux de construction, des jeux de cartes basés sur le calcul, des puzzles, lire, dessiner, travailler sur d'autres supports autonomes en lecture, etc. Pour recenser les enfants qui effectuent du travail en autonomie, une feuille était à leur disposition, et il leur a été demandé de s'inscrire avant d'effectuer chaque tâche. Ce recensement a permis de déterminer 6 périodes de travail en autonomie sur chaque semaine.

L'évaluation d'un jeu sérieux est souvent limitée à des informations basiques, essentiellement sur le temps passé sur le jeu, et si il a été fini ou pas[14]. Le jeu sérieux est alors considéré comme une boîte noire, en ne traitant que les données de session. Ici il a été souhaité traiter des informations plus précises sur l'utilisation des jeux, notamment pour chercher à comprendre et évaluer le comportement de l'enfant apprenant face aux jeux de QCM.

Pour cela, nous avons utilisé des "métriques de gameplay", qui sont des données quantitatives horodatées, directement produites par le code source d'un jeu, et rendant compte des différentes actions effectuées par un joueur dans le jeu (par exemple saut, déplacement, utilisation d'un objet, discussion avec un personnage, etc)[15]. Les métriques de gameplay permettent d'avoir une description exhaustive des interactions entre un joueur et un système de jeu, et permettent de simplifier l'analyse de l'expérience, en offrant un résumé de la session. Nous disposons alors de données quantitatives, comparables entre elles, et offrant une précision d'analyse plus fine qu'une observation complète réalisée pendant la session de jeu. De plus, une observation directe peut rajouter un biais, les joueurs n'interagissant pas de la même manière lorsqu'ils se savent observés par un spectateur extérieur.

Nous avons développé, en utilisant le moteur de jeux-vidéo Unity3D[16], un module de log automatique de métriques. Ces métriques sont enregistrées dans un fichier, avec l'identifant du joueur, la date précise (en milliseconde) de chaque action réalisée, et l'identifiant de cette action (avec des variables potentielles, comme par exemple sur la réussite d'une action true/false, ou la position en x et y d'un avatar, etc.). Ce module permet également d'enregistrer les métriques sur un serveur web, pour faire de la télémétrie, mais nous n'avons pas utilisé cette possibilité dans les expérimentations en classe.

Pour QCM, voici la liste non exhaustive des métriques que nous avons décidé de récupérer et de traiter dans cette étude. Chaque métrique étant horodatée, nous ne précisons pas cette information dans la liste suivante :

- **Général**

    Lancement du jeu (GAME_START, true)

    Fermeture du jeu (GAME_END, true)

    Pause (GAME_PAUSE, true/false)

- **Menu principal**

    Lancement de l'écran principal (MAIN_MENU_START, true)

    Sortie de l'écran principal (MAIN_MENU_QUIT, true)

- **Narratif**

    Lancement du jeu narratif (STORY_START, true)

    Entrée dans une nouvelle scène (STORY_SCENE_START, numéro)

    Sortie de la scène (STORY_SCENE_END, numéro)

    Ecran de fin (STORY_END_SEQUENCE, true)

- Labyrinthe

    Lancement du jeu labyrinthe (LABYRINTHE_START, true)

    Sortie du jeu labyrinthe (LABYRINTHE_END, true)

    Niveau courant (LABYRINTHE_LEVEL, numéro du niveau)

    Collision objets (LABYRINTHE_(KEY/BONUS/DOOR), true)

- Fiches

    Lancement de l'écran fiche (QUESTION_START, true)

    Écran exemple (QUESTION_EXAMPLE, lien vers l'image)

    Écran réponse (QUESTION_QCM, true)

    Réponse du joueur (QUESTION_ANSWER, numéro du choix, bonne/mauvaise réponse)

Les expériences menées en classe ont ainsi permis de récupérer environ 140000 lignes de métriques, qui ont été ensuite traitées de manière informatique pour obtenir des informations permettant de mettre en lumière[17] le comportement des apprenants. Les principaux résultats sont présentés dans la section suivante.

## 4 Résultats

Les résultats des expériences menées en classe sont présentés dans cette partie, en essayant de considérer trois grandes problématiques : la rapidité de prise en main des nouveaux supports par les élèves, l'influence des mécanismes de ludification sur la motivation des élèves et enfin la diversité des comportements par rapport à ces différents mécanismes.

L'évaluation des jeux sérieux en contexte scolaire est complexe et peut être appréhendée sous plusieurs angles. En effet, une multitude de méthodes d'évaluations existent, mais elles restent souvent propres à chaque jeu[18], ce qui explique la difficulté de proposer des méthodes génériques d'évaluation[19]. Généralement, l'évaluation est effectuée par rapport au temps passé sur une tâche d'apprentissage[20]. Dans les expériences présentées dans cet article, la tâche d'apprentissage intervient entre chaque jeu. Concernant l'aspect pédagogique, les résultats présentés sont donc essentiellement liés aux nombres de fiches effectuées, au temps passé sur ces tâches et au nombre d'erreurs.

### 4.1 Utilisation des jeux

Une première partie des observations concerne l'utilisation des jeux sérieux sur tablettes numériques, en regard de l'originalité du support et du manque d'habitude dans l'utilisation des tablettes numériques dans un contexte d'apprentissage. La première séance de chaque semaine a duré 30 minutes environ, tous les enfants jouant sur les tablettes jusqu'à ce que l'enseignant indique l'arrêt de l'activité. Au début, la plupart des enfants se montraient assez impatients de commencer, démontrant une grande curiosité. Cette motivation a été maintenue puisque lorsqu'il a été demandé oralement si les enfants souhaitaient continuer l'activité, tous sans exception ont répondu par l'affirmative avec enthousiasme. Le silence lors de chaque séance était très marqué, seulement troublé par des demandes d'ordre technique (par exemple, augmenter/diminuer le son) ou par la volonté de certains enfants de partager des réussites, en particulier sur le jeu ludique. Ce dernier point confirme l'intérêt d'envisager à l'avenir des expérimentations sur la dimension sociale des jeux (comparaison des scores par exemple).

Concernant la prise en main des tablettes, il a été remarqué une facilité d'utilisation, certainement appuyée par l'expérience des jeux sur tablettes pour la plupart des enfants à leur domicile. Certains avaient toutefois indiqué qu'ils n'utilisaient pas de tablette chez eux, sans pour autant montrer de difficulté particulière pour interagir. Les demandes d'aides étaient principalement liées à des réglages (volume, casque par exemple) ou à la gestion de rares problèmes d'exécution des jeux. Les enfants n'ont pas non plus semblé déstabilisés par les fiches de lecture, qu'ils n'avaient pourtant jamais utilisées auparavant.

Lors de la première séance sur tablette de chacun des deux groupes d'essai, les élèves ont testé les deux jeux : le jeu ludique et le jeu narratif. Ils ont majoritairement commencé par le jeu ludique, pourtant présent en seconde place sur le menu (voir le figure 4) : seulement 5 enfants sur 23 ont commencé par le jeu narratif. Il est à noter différents comportements : certains ont d'abord testé chacun des deux jeux avant d'en choisir un ; alors que d'autres ont préféré terminer l'un des deux avant de tester le suivant. Beaucoup ont également essayé le jeu de plateau lors des phases de travail en autonomie dans la semaine : au moins les deux tiers des enfants se sont montrés curieux pour cet autre jeu et l'ont essayé avec des camarades.

Les données d'utilisation permettent de mettre en avant quelques éléments importants. Tout d'abord, le nombre de fiches effectuées (c'est-à-dire traitées par les enfants, avec la bonne réponse) est de 418 pour les fiches au format papier, alors qu'il atteint 692 pour les fiches au format numérique. La différence est conséquente : 60% de plus pour les fiches numériques effectuées au travers des jeux sérieux. En moyenne, 26 fiches au format papier ont été traitées par chaque élève, pour 43 fiches au format numérique pour la même période de temps[3]. La figure 8 montre les variations du nombre de fiches uniques traitées par les différents élèves, selon les deux contextes. Les élèves les plus à l'aise en lecture ont terminé les deux niveaux de fiche quels que soient leurs formats, comme par exemple les élèves *A*, *F* ou *R*. En revanche, certains élèves qui n'ont pas traité beaucoup de fiches en papier, se montrent ensuite plus actifs, ce qui permet bien souvent de rattraper le retard pris la première semaine sur les fiches au format papier, comme par exemple les élèves *N*, *O* ou *Q*.

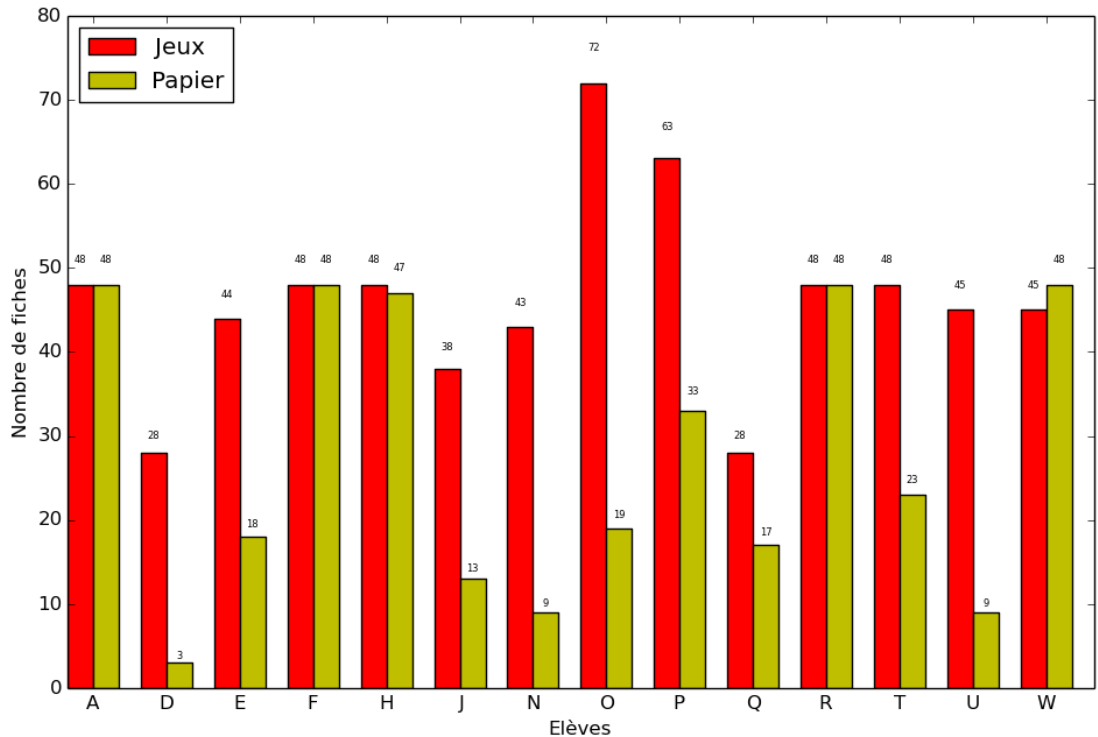

**Figure 8.** Variations du nombre de fiches uniques traitées par les différents élèves selon les deux contextes (papier et jeux). A cause d'erreurs de relevés d'activités et d'absences pouvant fausser les résultats, le groupe d'élèves a ici été réduit à seulement 14 élèves.

Cette différence sur le nombre de fiches traités selon le support peut s'expliquer en partie par la curiosité amenée par le support numérique. Cette justification peut ceci dit être tempérée, car les enfants n'avaient pas non plus manipulé les fiches au format papier auparavant. Une curiosité existait donc également pour ce format. De façon plus pratique, la correction des fiches au format papier par l'enseignante induit un déplacement dans la classe avec une attente potentielle, ralentissant alors le processus d'apprentissage pour les enfants travaillant sur support papier. De plus un seul exemplaire de chaque fiche est généralement disponible en classe au format papier, ce qui peut ralentir ou perturber certains enfants, qui ont parfois besoin de respecter un ordre établi, comme une routine d'organisation rassurante. Au final, les nombres de fiches traités par les enfants sur un même intervalle de temps appuient l'avantage pratique des fiches au format numérique sur les fiches au format papier.

Toutefois, il est indispensable d'étudier également l'impact de la ludification des fichiers d'apprentissage de la lecture, pour évaluer si l'augmentation du nombre de fiches traités est dû au passage des fiches au format numérique ou bien si la ludification a induit une motivation supplémentaire. En analysant les fiches abordées par les élèves, il apparaît que certains élèves ont effectué l'ensemble du fichier proposé, et, en continuant à jouer, ont été amené à refaire des fiches déjà traitées. Certains ont également choisi de reprendre les jeux sérieux pendant les périodes de travail en autonomie libre de la semaine et décidé de refaire des fiches déjà effectuées. De plus, aucun enfant (ayant pourtant terminé un fichier) n'a décidé d'arrêter de jouer. Sur l'échantillon étudié, 10 enfants sur le groupe de 23 ont terminé le fichier et ont néanmoins continué à jouer. Par exemple, l'élève *R* a traité plus de 140 fiches alors que le fichier n'en contient que 48. Il est intéressant de remarquer ici que cet élève s'était signalé comme ne jouant pas à la tablette chez lui. Pour ces cas, la motivation est donc clairement lié à la ludification plutôt qu'à la volonté de terminer le travail proposé.

---

[3]En prenant en compte les absences et erreurs de relevés, la comparaison entre fiche au format papier et au format numérique n'est effectué que sur un sous-ensemble de 14 enfants.

Un autre élément concernant les taux d'erreur mérite d'être mis en avant. Le nombre de mauvaises réponses aux fiches a été mesuré, et un taux d'erreurs moyen en a été déduit. La figure 9 représente le nombre de réponses justes et fausses pour chaque élève lors du travail sur les jeux sérieux. L'écart entre les taux d'erreurs pour les fiches au format numérique et ceux pour les fiches au format papier est important et clairement à l'avantage du format papier. En effet, le taux d'erreur sur format papier est de 4.5% alors qu'il atteint 22.5% sur format numérique. Plusieurs explications peuvent être avancées, mais la principale semble être liée à l'attention portée à la recherche des bonnes réponses. Le contexte ludique semble amener certains élèves à moins de rigueur dans la lecture. La différence de comportement est sur ce point assez marquée. L'élève $F$ par exemple, recensé parmi les élèves les plus à l'aise en lecture, a semblé bien moins concentré sur les jeux, en obtenant 25% d'erreurs alors qu'il n'en avait obtenu aucune sur les fiches au format papier. A l'opposé, certains élèves, moins nombreux, n'ont pas semblé être sensible au changement de contexte. Par exemple, l'élève $H$ a commis seulement 2 erreurs sur ses 133 fiches traitées (un seul fichier de 48 fiches, mais avec le traitement de la même fiche à plusieurs reprises). En travaillant sur les fiches papier, cet élève a effectué 4 erreurs sur les 48 fiches traitées, alors même que, pour cet élève, le fichier 3 traité lors des séances de jeux était plus compliqué que le fichier 2 traité sur feuille.

Ces deux enfants sont des élèves très sérieux, particulièrement soucieux de fournir un travail de qualité. Cette différence de comportement face à l'outil numérique - beaucoup plus d'erreurs que sur papier pour $F$ et très peu comme sur papier pour $H$ - peut trouver une explication dans la représentation de la tâche que chacun des deux élèves s'est construit. L'élève $F$, dont l'accès à la tablette chez lui est très contrôlé, avait formulé le fait que le "vrai travail" est celui sur format papier. Le format numérique représentait à ses yeux un "simple jeu" ne demandant donc pas d'implication particulière, selon ses propres critères. Au contraire, l'élève $H$ avait pleinement conscience d'être dans une situation d'apprentissage, dans un cas comme dans l'autre. Elle a donc appliqué la même rigueur, la même concentration sur format numérique ou papier.

Cette augmentation du taux d'erreurs, assez marquée pour certains élèves, mériteraient d'être étudiée plus en profondeur pour mettre en valeur l'ensemble des éléments qui peuvent perturber l'expérience de certains joueurs, comme par exemple la lisibilité des fiches sur tablettes.

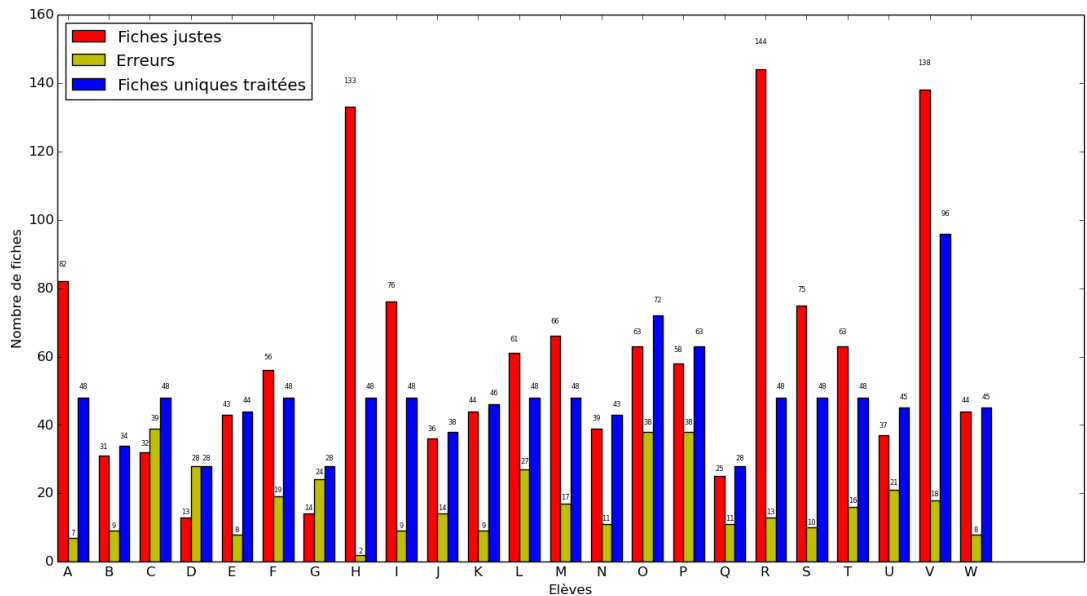

**Figure 9.** Variations du nombre de réponses correctes, du nombre d'erreurs et du nombre de fiches uniques traitées pour les différents élèves sur les fiches des jeux sérieux.

Malgré le nombre plus élevé d'erreurs lors du traitement des fiches dans les jeux sérieux, il apparaît qu'au final, le nombre total de fiches correctement effectuées par les élèves est plus important dans ce contexte : 692 fiches uniques ont été correctement traitées pour seulement 418 au format papier. Les élèves sont ainsi repassés sur les fiches fausses et y ont répondu de façon correcte. Une étude plus conséquente avec un nombre plus important d'élèves permettrait de comprendre l'origine des augmentations des taux d'erreurs chez certains élèves, et d'observer si une habitude d'utilisation lisserait ce taux d'erreurs, ainsi que le surplus observé de motivation. Il serait aussi intéressant de mesurer l'impact du travail effectué avec les jeux sur l'assimilation des stratégies de lecture. Les premières expériences présentées ici permettent cependant de mettre d'ores et déjà en avant l'intérêt de la ludification sur la motivation et mettent clairement en avant des différences de comportement entre les élèves dans un tel contexte.

### 4.2 Motivation

Dans cette partie, nous présentons des résultats sur la motivation induite par les jeux pour travailler l'apprentissage de la lecture en autonomie sur les tablettes. Comme indiqué dans le protocole présenté en section 3, les élèves, après une première séance encadrée de travail (jeux sérieux ou papier), avaient ensuite la possibilité de choisir les fiches de lecture lors de période de travail en autonomie, parmi d'autres activités habituelles (jeux de construction, jeux de cartes mathématiques, puzzles, lecture, dessin, ...). Selon leur groupe et la semaine, les élèves avaient soit accès aux fiches au format papier, soit aux jeux sérieux sur tablettes. Pour ces périodes d'autonomie, les élèves devaient s'inscrire en indiquant qu'ils allaient procéder à une activité en autonomie (sans indiquer laquelle). Les relevés automatiques d'activité sur tablettes permettent de connaître la proportion des élèves qui ont choisi de travailler en utilisant les jeux sérieux. L'expérimentation sur deux semaines permet d'observer l'évolution du comportement sur ces périodes d'autonomie, selon l'accès possible (ou non) aux jeux sérieux sur tablettes. Le tableau 1 récapitule le nombre de travaux effectués en autonomie pour chaque groupe, sur chaque semaine.

| Groupe | Semaine 1 | Semaine 2 |
|--------|-----------|-----------|
| 1      | **47**    | 39        |
| 2      | 30        | **48**    |

**Table 1.** Evolution du nombre de travaux effectués en autonomie dans la classe selon les groupes lors des deux semaines d'expérimentations.

Une augmentation significative (p-value de 0.038 en test $\chi^2$) du nombre d'activités en autonomie des élèves ayant accès aux jeux sérieux sur tablettes (le groupe 1 pour la semaine 1, le groupe 2 pour la semaine 2) par rapport aux élèves ayant accès aux fiches au format papier est observée. L'accès aux jeux sérieux sur tablettes semble donc induire une augmentation d'activités et de motivation. Il est important de noter que la curiosité qui pourrait entrer ici en jeu dans la modification des comportements touche également les fiches au format papier (non utilisées jusqu'alors par les élèves). Il faut évidemment pondérer les conclusions devant la courte durée de l'expérience présentée, le faible nombre d'élèves concernés ainsi que ce possible effet de curiosité. Une expérience sur une plus longue durée pourrait confirmer ou infirmer ces premières tendances.

Au niveau de la motivation, des différences de comportements entre les élèves peuvent être observées. Par exemple, l'élève *G* a effectué deux fois plus de travaux en autonomie en semaine 2 qu'en semaine 1, alors même qu'il est dans le groupe 1, c'est-à-dire ayant accès aux jeux sur la semaine 1. De même, l'élève *M* a réalisé 50% de plus de travaux en autonomie en semaine 1 qu'en semaine 2, alors qu'il avait accès aux jeux sur la semaine 2. Pour ces exemples, il est probable que les fiches au format papier les aient intéressés ou qu'ils n'aient plus envie de reprendre les jeux sérieux. A l'opposé, les élèves *E* ou *U* ont été beaucoup plus actifs sur les périodes d'autonomie lors de la semaine au cours de laquelle ils avaient accès aux jeux sérieux sur tablettes.

Ces observations semblent indiquer que, pour certains élèves, le format numérique ludifié des fiches d'apprentissage induit un surplus de motivation. Certains élèves sont moins sensibles à ce format. Il est toutefois intéressant d'étudier plus finement le comportement des élèves par rapport aux jeux, et analyser s'il y a également des différences de comportement selon les jeux, c'est-à-dire des sensibilités différentes aux dimensions de ludification choisies dans cette expérience. La section suivante détaille cette problématique.

### 4.3 Personnalisation

Dans cette section, différentes réactions d'élèves par rapport aux jeux proposés sont recensées. Il s'agit ici de démontrer les liens variés entre les joueurs et les jeux sérieux, selon les dimensions de ludification. La figure 10 permet de visualiser les différentes sessions de jeux pour chaque élève. Un nouveau segment indique une nouvelle session de jeu, alors qu'un changement de couleur indique un changement de jeu : trois couleurs différentes sont utilisées pour les trois jeux proposés : ludique, narratif et collaboratif (jeu de plateau collectif). La durée est représentée par la longueur du segment.

Tout d'abord, cette représentation permet de mettre en avant les différences de choix entre certains élèves. Par exemple, l'élève *K* n'a joué qu'une seule fois au jeu narratif. Il a terminé le jeu lors de la première séance mais n'a pas eu envie de recommencer le jeu. A la fin de la semaine, il a ainsi lancé 7 fois le jeu ludique mais une seule fois le jeu narratif. Cela peut s'expliquer par les interactions possibles avec les autres élèves, notamment pour obtenir un meilleur score (temps pour finir les niveaux) et ainsi se comparer à ses camarades. Ce mécanisme de motivation lié à la dimension sociale doit être expérimenté lors d'une prochaine expérience. Un autre exemple est l'élève *D* qui, au cours de la première séance, joue au jeu du labyrinthe, teste le jeu narratif sans le terminer, et préfère revenir au jeu ludique. Ces élèves, qui ont principalement rejoué au jeu du labyrinthe, sont donc plus sensibles à la dimension ludique.

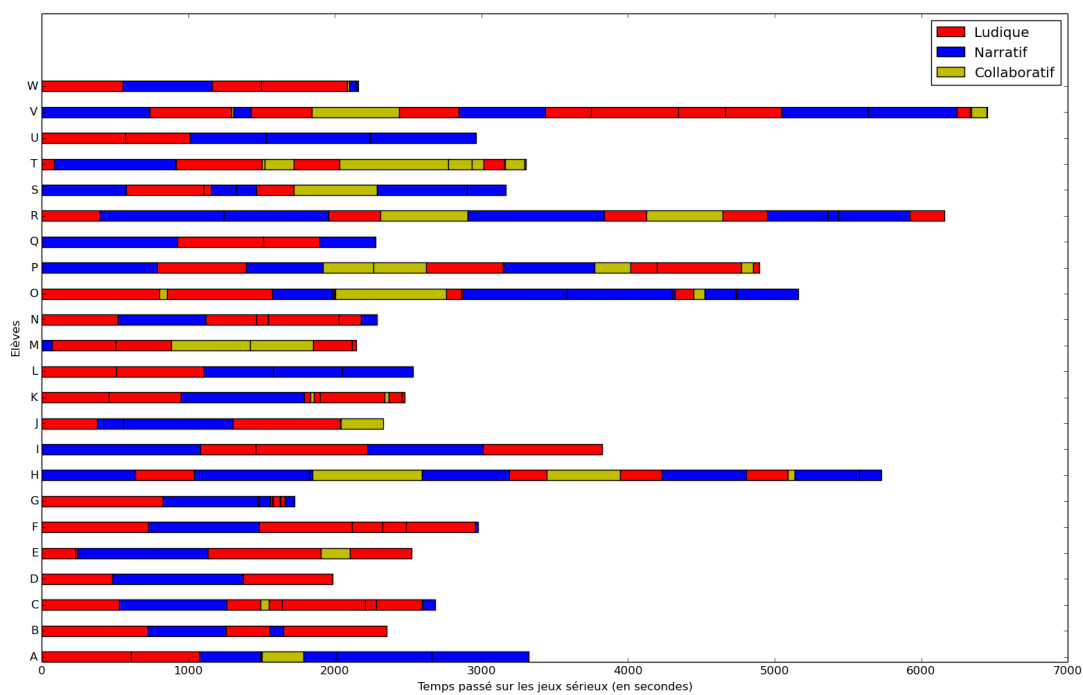

**Figure 10.** Visualisation des sessions de jeux au cours du temps pour chaque élève.

D'autres élèves effectuent des choix d'utilisation complètement différents, en fixant une attention particulière au jeu narratif. Par exemple, les élèves *A* et *H* ont lancé 8 fois le jeu narratif pour seulement 3 (respectivement 4) fois le jeu ludique du labyrinthe. Ainsi, l'élève *A* a commencé la première séance par le jeu ludique, l'a terminé, l'a recommencé avant de commencer le jeu narratif. A partir de ce moment-là, cet élève n'est plus revenu sur le jeu ludique et a ensuite essentiellement joué au jeu narratif. L'explication provient certainement du fait que cet élève *A* n'a pas eu le temps sur la première séance, puis au cours des séances en autonomie, d'aller au bout de l'histoire du jeu narratif. Il est envisageable que la motivation pour voir la fin de l'histoire ait été un facteur important pour le choix du jeu. L'élève *H* a lui alterné entre les jeux narratifs, ludiques et collaboratifs. Mais il est important de noter qu'à chaque fois qu'il a joué au jeu narratif, le temps consacré a été important, en progressant au sein de l'histoire à chaque session.

Dans le jeu ludique, un bonus (sous la forme d'un cornichon) apparaît dans les labyrinthes. Il n'a pas d'utilité pour le passage au niveau suivant, mais permet de débloquer un personnage (motivation de collection). La collection du cornichon déclenche une série de deux questions, comme pour la prise des clés. En analysant le comportement des élèves sur cet élément du jeu optionnel, il apparaît que le bonus a été pris 134 fois sur les 169 séquences durant lesquelles il a été disponible, ce qui représente un taux important de capture de cet élément optionnel de l'ordre de 80%. Ce taux est même sous-évalué, car une bonne partie des 20% restants sont essentiellement dus à un manque de temps pour terminer le niveau courant. Ce résultat est d'autant plus intéressant qu'au moment des expérimentations en classe, le mécanisme de collection n'était pas encore branché dans le jeu ludique : récupérer un cornichon ne débloquait pas de nouveaux personnages. Les élèves ont tout de même essayé de récupérer cet objet, certainement dans un souhait de terminer le niveau à 100%, et par la sémantique ludique attenante à cet élément de gameplay.

Une question sur la différence de comportements entre les filles et les garçons peut être soulevée. Par rapport à l'utilisation des jeux, il n'y a pas de différences significatives : 25% des élèves filles ont terminé le jeu narratif, pour 28% des garçons. Les élèves garçons ont été 79% à terminer au moins une fois le jeu ludique en comparaison des 50% de filles. Ce rapport légèrement plus élevé pour le jeu ludique pourrait indiquer une légère préférence des élèves garçons pour les jeux ludiques, explicable par le comportement observé pendant la première séance : certains élèves garçons semblaient en effet motivés à l'idée de comparer les scores obtenus sur le jeu ludique. Une expérimentation sur un plus grand nombre d'élève permettrait de confirmer ou d'informer cette hypothèse.

Sur la première séance de jeu, tous les élèves ont effectué entre 2 et 3 sessions de jeux. Certains ont terminé un jeu avant de commencer le suivant, tandis que d'autres ont d'abord testé les deux jeux avant d'essayer d'en terminer un. A la fin des séances, 6 élèves ont au moins une fois terminé le jeu narratif, alors que 15 élèves ont fini au moins une fois les 4 niveaux du jeu ludique. Seuls 4 élèves ont fini les deux jeux : *E*, *H*, *P* et *R*. Une observation essentiel est la différence de comportement devant les 2 différents jeux, puisque certains ont été motivés pour terminer l'un des deux jeux, mais le second. De plus, parmi les 4 élèves ayant terminé les deux jeux, on retrouve deux élèves *E* et *P*, qui ne sont pas parmi les élèves les plus à l'aise en lecture. La ludification des fiches a semblé les motiver. Cependant, pour l'élève *P*, cette motivation s'est traduite par un surplus d'erreurs. Pour lui, l'envie d'avancer a semblé prendre le pas sur la concentration et la qualité des réponses.

Comme indiqué dans la section 3, de nombreuses données ont été obtenues depuis l'utilisation des jeux par les élèves : interactions, temps passé, fiches traitées, etc. Ces différentes données peuvent permettre d'obtenir, après analyses précises, des éléments importants sur le plan pédagogique, que ce soit au niveau de la classe ou au niveau de chaque élève. Au niveau de la classe, l'analyse des données permet par exemple d'avoir une vision précise des compétences acquises ou alors des difficultés rencontrées sur certaines questions. Ceci peut traduire soit un besoin collectif d'approfondir certaines compétences avec les fiches concernées, soit une difficulté de compréhension lors de la lecture de ces fiches (formulation des phrases, des consignes, clarté de la tâche à accomplir). Pour illustrer cela, toutes les réponses des élèves ont été compilées pour établir un taux d'erreur par fiche. Ce taux d'erreur pour les 96 fiches utilisées est représenté par la figure 11. Des variations importantes du taux d'erreur en fonction des fiches peuvent être observées. Ces variations ne sont pas linéaires en fonction de la difficulté des fiches, ce qui peut révéler des difficultés spécifiques d'apprentissage. Par exemple, il est observé pour la question 49 un taux d'erreur assez important (le troisième taux le plus important). Sur cette question, l'élève a le choix entre 4 phrases dont certaines lettres sont masquées. L'analyse des réponses fausses est particulièrement intéressante : la réponse attendue est *c'est un gros livre* alors que l'analyse des erreurs indique que 7 enfants sur 8 qui ont fait une erreur ont répondu *c'est un grand livre*. Par rapport aux autres réponses et au regard des lettres cachées, il semble ici que la difficulté provient des deux lettres muettes : *s* à la fin du mot *gros* et *d* à la fin du mot *grand*. Cette capacité à comprendre les lettres muettes (en cherchant un mot de la même famille dans lequel cette lettre s'entend : gros/grosse, grand/grande) est une compétence travaillée jusqu'à la fin du cyle 2, et reste donc une difficulté récurrente en Cours Préparatoire, puisqu'en cours d'acquisition. L'analyse des erreurs semble cohérente, donnant un élément important pouvant être pris en compte sur le plan pédagogique.

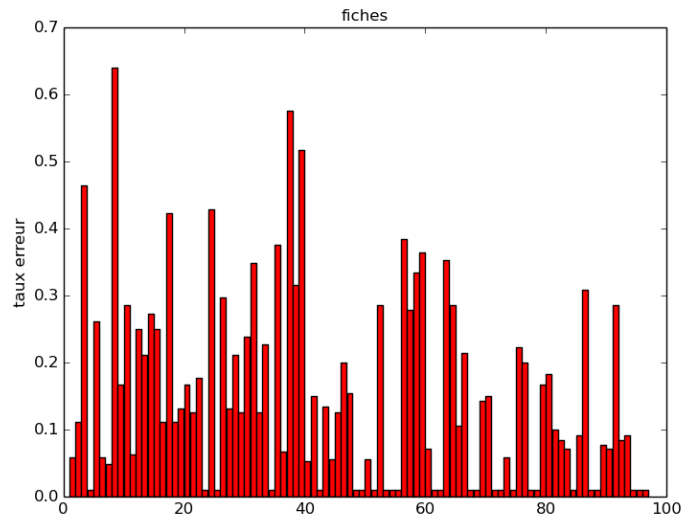

**Figure 11.** Variations du taux d'erreur pour les différentes fiches pédagogiques.

Les analyses de métriques obtenues pour chaque élève permettent également de mettre en relief d'éventuelles difficultés par rapport au groupe. Par exemple, si on étudie les temps passés à la résolution des questions pour chaque élève, des différences importantes sont observées : sur la première séance, le rapport est de 3 entre l'élève *H* le plus rapide (temps médian de 33 secondes) et l'élève *Q* le plus lent (temps médian de 10 secondes). Ainsi, en mettant directement en relation avec les taux d'erreur, il est possible d'obtenir la figure 12, qui propose une visualisation du taux de réussite en fonction du temps médian passé sur la résolution des questions des fiches. Deux couleurs différentes ont été utilisées pour différencier les deux groupes d'élèves et s'assurer qu'il n'y ait pas de biais introduit par les différences de conditions expérimentales. La largeur du point est proportionnelle au nombre de fiches traitées par l'élève : plus le point est large, plus le nombre de fiches traités par l'élève est important, et donc plus la donnée est fiable. En haut à gauche peuvent donc être visualisés les élèves qui répondent rapidement et correctement, alors qu'en bas apparaissent les élèves plus en difficulté sur ces questions de lecture. Ainsi, l'élève *H* est un exemple d'élève à l'aise avec les questions de lecture, et qui y répond dans la majorité des cas à la fois correctement et rapidement. Mais cette analyse permet de mettre en relief un comportement particulier d'élèves. Par exemple, il peut être utile de conseiller aux élèves *C*, *D*, *G* et *P* de prendre un peu plus de temps pour répondre aux fiches, éventuellement aussi aux élèves *M*, *S* ou *V*, qui ont certes un bon taux de réussite, mais qui pourraient peut-être faire mieux en prenant plus de temps de réflexion. L'élève *P*, peu en confiance en lecture, a habituellement un rythme de travail plutôt lent mais pertinent. Ces résultats indiquent donc un changement de comportement probablement induit par le changement de support et la ludification. De même, l'élève *V*, un des meilleurs lecteurs de la classe, sérieux et appliqué dans le travail pour ne pas faire d'erreur, a visiblement été dérangé par la ludification. La ludification a provoqué chez lui une accélération de son rythme de travail et il a donc fait plus d'erreurs qu'on ne pouvait l'attendre. Une expérimentation plus longue permettrait de voir si une plus grande habitude des jeux sérieux atténuerait cet effet auprès de lui. Faire le même type d'analyse pour chaque élève permettrait aussi de voir plus en détails si certaines questions ont nécessité plus de temps, ou si la concentration a été moindre sur certaines périodes de temps, entraînant un temps plus court de réponse mais un nombre accru d'erreurs. Des conseils personnalisés peuvent ainsi être directement déduits de ces analyses.

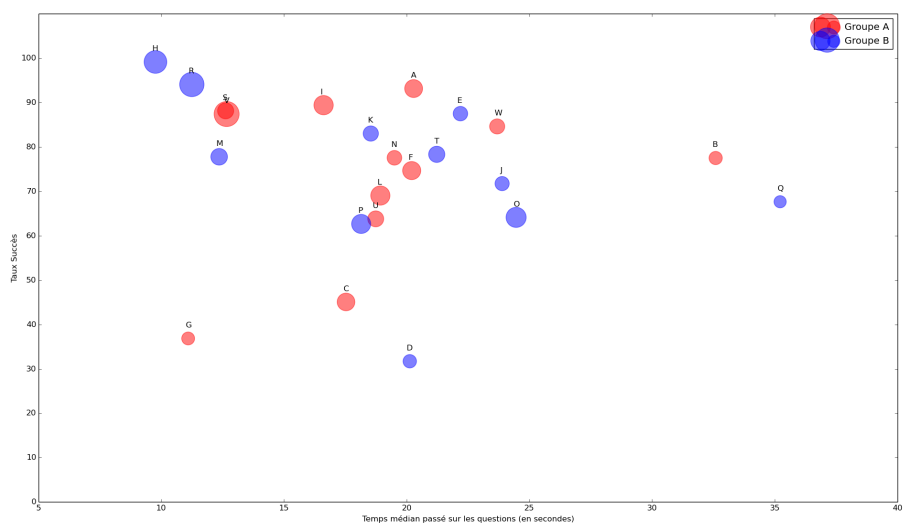

**Figure 12.** Variations du taux de réussite en fonction du temps médian passé à répondre aux questions des fiches pédagogiques. La largeur du point est proportionnelle au nombre de fiches traitées par l'élève et deux couleurs différentes permettent de visualiser les deux groupes différents d'élèves.

## 5 Conclusions et Perspectives

Deux mécanismes de ludification reposant sur les dimensions ludique et narrative ont été expérimentés dans une classe primaire. Les résultats sont encourageants. S'ils sont encore exploratoires, ils ouvrent la voie a de futurs travaux permettant d'en étudier la validité à plus grande échelle (effectif, nombre de fiches, durées des sessions, etc).

Les points principaux mis en valeur par résultats obtenus concernent différences de comportement entre élèves face aux jeux sérieux : certains sont perturbés par la ludification, certains y trouvent une motivation, alors que certains seront particulièrement sensibles à une dimension unique de ludification, etc. Ces observations appuient le besoin de ne pas proposer un seul type de jeux sérieux, mais de diversifier les mécanismes de ludification pour que chaque élève y trouve son intérêt en lien avec son style d'apprentissage. Ainsi, un exercice précis avec objectif pédagogique, comme les fiches d'apprentissage en autonomie de la lecture dans les études présentées ici, pourrait être ludifié de différentes manières, en appuyant sur différents mécanismes d'immersion. Les mécanismes expérimentés dans cet article reposent seulement sur les dimensions ludiques et narratives, mais il faudrait étendre à d'autres dimensions (sociale, kinéstésique, émotionnelle, ou stratégique).

Les nouveaux besoins induits par ces multiplications des ludifications seraient l'automatisation de la recherche de jeux, et surtout le recommandation des différents jeux, en fonction du profil de l'apprenant et/ou des objectifs pédagogiques de l'enseignant. Il faudrait d'une part être capable de maîtriser ces mécanismes pour proposer différents types de jeux centrés sur la même problématique pédagogique. Mais il faudrait également être capable de guider l'élève et l'enseignant devant le nombre de jeux et leurs différents types, et aller ainsi vers des moteurs de recommandation adaptés et personnalisés. Il s'agirait donc ici de développer des moteurs de recommandation combinant à la fois les aspects de ludifications, mais également les aspects pédagogiques.

Par ailleurs, d'autres perspectives novatrices peuvent être envisagées, notamment liées au *learning analytics*, c'est-à-dire le traitement des données issues des logs d'utilisation, permettant d'envisager l'amélioration de la personnalisation, la validation automatique des acquis ou encore l'aide à la décision pour le choix des apprentissages ludifiés à considérer[21]. La génération automatique de jeux à partir de supports pédagogiques numériques, en sa basant sur plusieurs mécanismes de ludification est également une voie à considérer.